\documentclass[runningheads]{llncs}
\usepackage[T1]{fontenc}
\usepackage{float}
\usepackage{placeins}

\usepackage{graphicx}
\usepackage{amsmath}
\usepackage{amssymb}
\begin{document}
\title{Aggregating User Preferences while Ensuring Equity, Diversity, and Inclusion using Graph Summarization\thanks{An interactive dashboard with the full experimental results is available at \texttt{https://sitaacisse-boop.github.io/EDI/}.}}

\titlerunning{Aggregating Preferences with EDI via Graph Summarization} % If the paper title is too long for the running head, you can set
\author{Adji Marieme Sita Cissé\inst{1,2} \and Malek Mouhoub\inst{2} }
\authorrunning{A.~M.~S.~Cissé}
\institute{Université Paris-Saclay, UVSQ, France
\and
Department of Computer Science, University of Regina, Regina, SK, Canada\\
\email{louzsita29@gmail.com, Malek.mouhoub@uregina.ca}\\
}

\maketitle              % typeset the header of the contribution
\begin{abstract}
Aggregating the preferences of diverse user groups into a collective outcome
raises fundamental challenges of equity, diversity, and inclusion (EDI):
classical aggregation rules such as Borda and Condorcet have no mechanism
to prevent results from systematically favoring majority groups, collapsing
onto homogeneous items, or under-representing minorities. These tensions
surface concretely in group recommendation, where optimizing for utility
alone entrenches majority bias and filter bubbles. We address this problem
through EDI-constrained graph summarization. User preferences are modeled
as a weighted attributed bipartite graph, and a greedy coarsening algorithm
iteratively merges user nodes while enforcing three structural EDI criteria:
an equity gap constraint ($\Delta E$), an intra-list diversity constraint
(ILD), and a group inclusion constraint. Rather than correcting fairness
after aggregation, our method embeds EDI preservation directly into the
graph structure.

We evaluate across five datasets spanning four domains --- movie
recommendation (MovieLens 100k and 1M), social dating (libimseti.cz),
educational rating (Rate My Professors), and academic authorship in
AI/ML/CS (OpenAlex, 2018--2023). No single baseline simultaneously
satisfies all three EDI criteria across domains. Our method, AURORA,
achieves the largest and most consistent diversity gains over classical
voting rules (Borda, Weighted Borda, Condorcet), and on MovieLens 100k at
$k=20$ it simultaneously improves all three EDI criteria over both Borda
and Condorcet. On item-side equity settings, it uniquely combines
high diversity (ILD\,$=$\,0.808) with the highest female item
representation ($\mathrm{frac}_F=60\%$) on Rate My Professors --- though
at a moderate equity cost relative to Borda on this dataset --- and
achieves the lowest equity gap ($\Delta E = 0.031$) on OpenAlex, where
Borda-based methods recommend zero female authors. On libimseti.cz, the
only dataset where the sensitive attribute is present on both sides of the
bipartite graph, our method --- like classical voting rules --- does not
reduce the equity gap, a limitation we
analyze and connect to prior findings that demographic parity is not
always an appropriate target when preferences genuinely differ across
groups. These results demonstrate that embedding EDI constraints into
aggregation structure yields more robust fairness--diversity trade-offs
than post-hoc approaches, while also revealing that no single
compression budget or structural mechanism resolves the equity--diversity
tension uniformly across domains.
\end{abstract}
\section{Introduction}

As algorithmic systems increasingly mediate access to information, opportunities,
and resources, ensuring that their outcomes do not systematically disadvantage
certain groups has become a central concern across many domains, from hiring and
lending to content curation~\cite{mehrabi2021survey}. Equity, diversity, and
inclusion (EDI) capture three complementary facets of this concern: equity asks
that outcomes not systematically favor one group over another, diversity asks that
outcomes not collapse onto a narrow set of options, and inclusion asks that
minority groups be adequately represented in the outcome. Recommender systems
are a paradigmatic case in which these tensions surface concretely. They have
become essential tools for navigating the vast amount of content available online,
from movies to products to services, and their traditional goal is to maximize
utility --- that is, to predict as accurately as possible the items a user will
like. However, optimizing for utility alone has well-documented drawbacks. It
tends to reinforce the so-called Matthew effect, in which already-popular items
are recommended ever more often, and it can trap users in filter bubbles of
homogeneous content. These effects raise EDI concerns: collective recommendations
may systematically favor the tastes of majority groups, offer little variety, and
under-represent minorities.

Addressing EDI in recommendation requires aggregating the preferences of many
users into a collective outcome while treating user groups fairly. Existing
approaches typically enforce fairness by modifying the input
preferences~\cite{kamiran2012data}, by modifying the output ranking through
re-ranking~\cite{singh2018fairness}, or by regularizing the learning
algorithm~\cite{zhao_fairness_2024}. In parallel, graph summarization offers a
powerful way to reduce large user--item preference graphs into compact, tractable
representations, but it has traditionally been guided by structural fidelity,
storage, or query efficiency, never by the preservation of fairness. To the best
of our knowledge, no existing work makes the preservation of equity, diversity,
and inclusion an explicit criterion when summarizing a preference
graph~\cite{liu_graph_2018}.

In this paper, we address this gap. We represent user preferences as a
weighted attributed bipartite graph and propose AURORA (\textit{Attributed Unified gRaph cOarsening
for equitable RecommendAtion}), an approach that aggregates
preferences through graph summarization under EDI constraints: instead of merging nodes solely
to minimize structural error, we constrain the summarization so as to bound
the degradation of equity, diversity, and inclusion. Our main objective is
to construct a summary that reduces the size of the preference graph while
preserving these three properties, and to compare it against five classical
aggregation rules: Average Score, Borda, Weighted Borda, Condorcet, and
Fair Re-rank. We instantiate and evaluate our approach across five datasets
spanning four domains --- movie recommendation (MovieLens 100k~\cite{harper2015movielens}
and 1M), social dating (libimseti.cz), educational rating (Rate My
Professors), and academic authorship in AI/ML/CS (OpenAlex, 2018--2023)
--- using gender as the sensitive attribute in all cases, applied either
to users (user-side equity) or to recommended items (item-side equity).

The remainder of this paper is organized as follows. Section~2 introduces
the necessary background. Section~3 reviews related work. Section~4 presents
our proposed methodology. Section~5 reports the experimental evaluation,
and Section~6 concludes.

\section{Background}
\label{sec:background}

This section introduces the concepts and notation used throughout the paper:
the preference graph, preference aggregation, graph summarization, and the
notions of equity, diversity, and inclusion.

\subsection{Preference Graph}

We represent preference data as a weighted attributed bipartite graph
\[
  G = (U \cup I,\; E,\; w,\; s),
\]
where $U = \{u_1, \ldots, u_m\}$ is the set of user nodes, $I = \{i_1,
\ldots, i_n\}$ is the set of item nodes (e.g.\ movies, professors, or
researchers), and $E \subseteq U \times I$ is the set of edges. An edge
$(u, i) \in E$ exists if and only if user $u$ has rated item $i$. The weight
function $w : E \rightarrow [1, 5]$ gives the rating assigned by $u$ to $i$.
The sensitive attribute $s$ is defined on user nodes in \emph{user-side
equity} settings ($s : U \rightarrow \{F, M\}$, e.g.\ MovieLens, libimseti)
or on item nodes in \emph{item-side equity} settings ($s : I \rightarrow
\{F, M\}$, e.g.\ Rate My Professors, OpenAlex), where $\{F, M\}$ denotes
the two gender groups. The graph is bipartite: edges connect only users to
items, never two users or two items.

\subsection{Preference Aggregation}

Preference aggregation is the task of combining the individual preferences
of many users into a single collective outcome, typically a ranking or a
top-$k$ set of items~\cite{ricci2011recommender}. Classical aggregation
rules provide our baselines. The \emph{average score} ranks items by their
mean rating; the \emph{Borda count} converts each user's ratings into ranks
and sums them; \emph{Weighted Borda} normalizes Borda scores by group size
to compensate for demographic imbalances; and the \emph{Condorcet} method
compares items pairwise, an item winning if a majority of users prefer it.
None of these rules explicitly controls EDI during aggregation, which
motivates the approach proposed in this paper.

\subsection{Graph Summarization}

Graph summarization reduces a large graph to a smaller one that preserves
selected properties of the original. We focus on the grouping (or
aggregation) family in which nodes are merged into \emph{supernodes}
according to their similarity and the edges between groups are aggregated
into \emph{superedges}~\cite{liu_graph_2018}. The process of merging nodes
to obtain a smaller graph is called \emph{coarsening}. The resulting summary
is smaller and faster to process, but the merging necessarily discards
information; the central question is therefore which properties the summary
should preserve.

\subsection{Equity, Diversity, and Inclusion}

We consider three distinct notions, jointly referred to as EDI. \emph{Equity}
refers to the fair treatment of groups defined by the sensitive attribute,
so that no group is systematically favored or disadvantaged by the collective
outcome --- whether the group is defined on users (user-side equity) or on
recommended items (item-side equity). \emph{Diversity} refers to the variety
of the recommended items, avoiding lists dominated by a single type or by
the most popular items. \emph{Inclusion} refers to ensuring that minority
groups are represented in the outcome at least in proportion to their size.
These three notions are distinct and can be in tension: improving one may
degrade another. Our work treats them as properties to be preserved when
summarizing the preference graph, and the precise metrics used to quantify
them are defined in Section~4.

\section{Literature Review}\label{sec:litreview} Our work sits at the intersection of four research areas: preference
aggregation and social choice, graph summarization, algorithmic fairness
(equity, diversity and inclusion), and fairness-aware recommender systems.
We review each in turn, and then identify the gap that our contribution
addresses.

\subsection{Preference Aggregation and Social Choice}

Preference aggregation, the problem of combining the individual preferences
of many users into a single collective outcome, is the theoretical foundation
of our work and is studied within computational social choice theory. Roy et
al.~\cite{roy_fairness_2024} structure the problem along three dimensions,
input format, output format, and aggregation rule, and survey how fairness
has been addressed across classical voting rules such as Borda, Copeland, and
single transferable vote. A central result framing the field is Arrow's
impossibility theorem, which shows that no aggregation rule can simultaneously
satisfy a small set of reasonable fairness criteria; this motivates reasoning
in terms of \emph{preserved properties} rather than a single optimal rule.
Notably, Roy et al.~\cite{roy_fairness_2024} distinguish between making an
outcome fair by modifying the input preferences and by modifying the output
ranking, an opposition we return to when positioning our own approach.
Building on this tradition, Kellerhals and Peters~\cite{kellerhals_proportional_2024}
connect proportional clustering to multiwinner voting and show that three fairness notions, proportional fairness, individual fairness, and the transferable core, are closely related and can be satisfied simultaneously. Crucially for our setting, their analysis operates on a graph metric, grouping nodes into fair clusters, which is conceptually close to summarizing a preference graph while preserving fairness guarantees.

\subsection{Graph Summarization}

Graph summarization aims to condense a large graph into a smaller
representation that preserves selected properties of the original, thereby
reducing storage, accelerating computation, and improving interpretability.
Liu et al.~\cite{liu_graph_2018} provide a reference survey of the field and
organize existing methods into four families: grouping (aggregating nodes
into supernodes based on structural or attribute similarity), bit compression,
simplification or sparsification (removing less important nodes or edges), and
influence-based approaches. A central observation of their survey is that
there is no single definition of a \emph{good} summary: the quality of a
summary depends on the objective it is meant to serve. They further
distinguish between plain, labeled (attributed), and dynamic graphs, a
distinction directly relevant to our setting, since our preference graph
carries a sensitive attribute on user nodes and is therefore a labeled graph.
More recent work, surveyed by Shabani et
al.~\cite{shabani_comprehensive_2024}, applies graph neural networks to
summarization and classifies approaches into aggregation, selection, and
transformation. Notably, this survey identifies \emph{task-based}
summarization, producing summaries tailored to a specific downstream need
rather than generic ones, as an open research direction, and highlights the
difficulty of defining quality metrics that depend on the target application.
Our work belongs to the grouping/aggregation family and directly addresses
this open direction: we propose a summary whose quality criterion is the
preservation of equity, diversity, and inclusion.

\subsection{EDI in Recommender Systems and Algorithmic Fairness}

The fairness, diversity, and inclusion (EDI) dimension is central to our work,
and each notion has its own formalizations. Yao and
Huang~\cite{yao_new_2017} study fairness in collaborative filtering and show
that demographic parity is often inappropriate for recommendation, since user
preferences genuinely depend on sensitive attributes such as gender; instead,
they propose four unfairness metrics (value, absolute, underestimation, and
overestimation unfairness), evaluated on MovieLens. We adopt their value
unfairness as the basis for our equity metric. Leonhardt et
al.~\cite{leonhardt_user_2018} focus on user-side fairness and provide
empirical evidence of a direct trade-off between diversity and user fairness:
increasing aggregate diversity substantially increases disparity among users,
as measured by two Gini-based metrics. This confirms that EDI objectives are
in tension and must be balanced rather than jointly maximized. Zhao et
al.~\cite{zhao_fairness_2024} survey fairness and diversity in recommender
systems, treat them jointly rather than in isolation, and classify mitigation
techniques into pre-processing, in-processing, and post-processing; they also
frame the problem as multi-objective optimization over a Pareto frontier of
accuracy, fairness, and diversity. Finally, Dong et
al.~\cite{dong_fairness_2023} survey fairness specifically in graph mining,
the closest area to AURORA. They highlight that graphs raise fairness challenges
absent from i.i.d. data, notably structural bias arising from homophily, and
classify fairness notions (group, individual, counterfactual, degree-related,
and application-specific) as well as mitigation techniques. Among the latter,
edge rewiring is the only one that acts on graph structure, as AURORA does;
however, it modifies the original graph by adding or removing edges, whereas
we reduce the graph by merging nodes. Our equity metric corresponds to group
fairness in their taxonomy.

\subsection{Social Choice for Fairness-Aware Recommendation}

The work closest to AURORA in spirit combines social choice and fairness
directly within recommendation. Aird et al.~\cite{aird_exploring_2023}
introduce SCRUF-D, an architecture in which each fairness concern is modeled
as an independent agent that expresses preferences over items; these agents
vote alongside the base recommender, through voting rules such as Borda,
Copeland, and Ranked Pairs, to produce the final ranking. A follow-up
study~\cite{aird_social_2024} extends this framework to heterogeneous
fairness, handling several fairness notions that differ in logic and
measurement within a single system, including group proportional fairness,
and evaluates it on MovieLens. Both works confirm a recurring finding:
fairness and accuracy are in tension, and different aggregation rules trade
them off differently. This line of work is our most direct conceptual
competitor, yet it differs from AURORA in a fundamental way: SCRUF-D acts at the
moment of producing the final ranking, that is, as a post-processing or
re-ranking step, whereas we act upstream on the structure of the summarized
graph. Their thorough exploration of the post-processing route leaves the
structural route open. We do not implement it as an experimental baseline:
SCRUF-D requires a per-fairness-concern utility function and an inter-agent
negotiation mechanism (Borda, Copeland, or Ranked Pairs voting among agents)
that would need to be designed and calibrated separately for each of our four
corpora --- a methodological contribution in its own right, distinct from our
fixed-EDI-metric evaluation protocol. Fair Re-rank, our post-processing
baseline, remains comparable in spirit (adjusting the final ranking under a
fairness constraint) without requiring this multi-agent architecture; a
faithful SCRUF-D implementation is left as a priority baseline for future
work.

\subsection{Research Gap}

Across these four areas, a consistent picture emerges. Fairness in
recommendation has been addressed mainly through input or output modification,
through pre-, in-, or post-processing, or through edge rewiring on the graph
structure; graph summarization, in turn, has traditionally optimized
structural fidelity, storage, query speed, or denoising. To the best of our
knowledge, no existing work makes the preservation of equity, diversity, and
inclusion an explicit quality criterion of a graph summary. This is precisely
the gap our work addresses. We therefore study the following research
question: given a weighted bipartite preference graph, how can we construct a
summary, through community detection and coarsening, that reduces its size
while bounding the degradation of equity, diversity, and inclusion, compared
to classical aggregation rules?

\section{Proposed Methodology}\label{sec:methodology}

Building on the graph model and background introduced in Section~2, this section
defines the three EDI metrics precisely, formulates the constrained summarization
objective, and describes how we derive a collective ranking from the summarized
graph.

\subsection{EDI Metrics}

Let $R = (r_1, \ldots, r_k)$ denote a collective top-$k$ ranking produced from
the preference graph, and let $\mathcal{G}_F, \mathcal{G}_M \subset U$ be the
two user groups induced by the sensitive attribute $s$.

\smallskip\noindent\textbf{Equity.}
We measure equity through the value unfairness metric of Yao and
Huang~\cite{yao_new_2017}. For each group $g \in \{F, M\}$, the group utility
over the top-$k$ list is
\begin{equation}
  \mathrm{utility}_g(R)
    = \frac{1}{|\mathcal{G}_g|}
      \sum_{u \in \mathcal{G}_g}
      \frac{1}{k} \sum_{i \in R} w(u, i),
\end{equation}
where $w(u,i)=0$ when $(u,i)\notin E$. The equity gap is then
\begin{equation}
  \Delta E(R) = \bigl|\mathrm{utility}_F(R) - \mathrm{utility}_M(R)\bigr|.
\end{equation}
A value of $\Delta E$ close to zero indicates that both groups derive similar
average satisfaction from the collective outcome.

\smallskip\noindent\textbf{Diversity.}
We measure diversity through the intra-list diversity (ILD) metric:
\begin{equation}
  \mathrm{ILD}(R) = \frac{2}{k(k-1)}
    \sum_{\substack{i,j \in R \\ i \neq j}} \mathrm{dist}(i,j),
\end{equation}
where $\mathrm{dist}(i,j) = 1 - \mathrm{cosine}(\mathbf{v}_i, \mathbf{v}_j)$
and $\mathbf{v}_i \in \mathbb{R}^m$ is the rating vector of item $i$ across all
users, with $\mathbf{v}_i[u] = w(u,i)$ if $(u,i)\in E$ and $0$ otherwise.
A higher ILD indicates a more varied recommendation list.

\smallskip\noindent\textbf{Inclusion.}
We measure inclusion through a proportional fairness criterion grounded in the
core concept of Kellerhals and Peters~\cite{kellerhals_proportional_2024} and
operationalized as the group proportional fairness metric of Aird et
al.~\cite{aird_social_2024}. Let $\theta$ be a relevance threshold and
$\alpha_g = |\mathcal{G}_g|/|U|$ the population share of group $g$. The
inclusion score of group $g$ is
\begin{equation}
  \mathrm{inclusion}_g(R)
    = \frac{1}{|\mathcal{G}_g|}
      \sum_{u \in \mathcal{G}_g}
      \frac{|\{i \in R : w(u,i) \geq \theta\}|}{k}.
\end{equation}
Inclusion is satisfied for group $g$ when $\mathrm{inclusion}_g(R) \geq
\alpha_g$: a group representing $\alpha_g$ of the user population should find
at least the same proportion of top-$k$ items relevant. We set $\theta = 4.0$
as our default on the 1--5 scale and compare with $\theta = 3.5$, the
threshold used by Aird et al.~\cite{aird_social_2024} to define a ``liked''
item on MovieLens (rating $>3$); this threshold is recalibrated per dataset
to preserve the same semantics across different rating scales (see
Section~\ref{sec:experimentation}).

\paragraph{Item-side equity.}
In item-side equity settings (Rate My Professors, OpenAlex), the sensitive
attribute is defined on items rather than users. The primary inclusion metric
is then the fraction of female items in the top-$k$ list:
\[
  \mathrm{frac}_F(R) = \frac{|\{i \in R : s(i) = F\}|}{k}.
\]
The equity gap $\Delta E$ is adapted accordingly, comparing the average
utility received by female and male items across all users in the graph.

\subsection{EDI-Constrained Graph Summarization}
Rather than enforcing EDI as a post-hoc correction on an already-computed
ranking, our approach embeds EDI preservation directly into the coarsening
process. Each candidate merge of user nodes is accepted only if the resulting
summarized graph maintains the equity gap, intra-list diversity, and group
inclusion of the original graph within prescribed tolerances. This structural
integration ensures that fairness is not traded off against compression after
the fact, but is preserved as the graph is progressively reduced.

\smallskip\noindent\textbf{Coarsening.}
Starting from $G = (U \cup I, E, w, s)$, we iteratively merge pairs of user
nodes into supernodes based on the similarity of their rating profiles. The
result is a summarized graph $G' = (U' \cup I, E', w', s')$ with $|U'| < |U|$,
where each supernode $S \subseteq U$ represents a group of original users.
The aggregated weight of supernode $S$ toward item $i$ is
\begin{equation}
  w'(S, i) =
    \frac{\displaystyle\sum_{u \in S:\,(u,i)\in E} w(u,i)}
         {|\{u \in S : (u,i) \in E\}|},
\end{equation}
when at least one member of $S$ has rated $i$. The collective top-$k$ ranking
is then derived from $G'$ using the aggregation rules described in Section~2.2.
The number of accepted merges is bounded by a merge budget expressed as a
fraction of $|U|$ rather than an absolute count, so that the same relative
compression target applies regardless of dataset size; Section~5 reports a
robustness sweep over this budget.

\smallskip\noindent\textbf{EDI preservation constraints.}
A merge producing candidate summary $G'$ is accepted if and only if
\begin{align}
  \Delta E(G') &\leq \Delta E(G) + \varepsilon_E, \\
  \mathrm{ILD}(G') &\geq \mathrm{ILD}(G) - \varepsilon_D, \\
  \mathrm{inclusion}_g(G') &\geq \mathrm{inclusion}_g(G) - \varepsilon_I
  \quad \forall\, g,
\end{align}
where $\varepsilon_E, \varepsilon_D, \varepsilon_I \geq 0$ are degradation
tolerances controlling the trade-off between compression and EDI preservation.

\smallskip\noindent\textbf{Formal objective.}
The overall problem is
\begin{equation}
\begin{aligned}
  \min_{G'} \quad & \mathcal{L}(G, G') \\
  \text{s.t.} \quad
    & \Delta E(G') \leq \Delta E(G) + \varepsilon_E, \\
    & \mathrm{ILD}(G') \geq \mathrm{ILD}(G) - \varepsilon_D, \\
    & \mathrm{inclusion}_g(G') \geq \mathrm{inclusion}_g(G) - \varepsilon_I
      \quad \forall\, g,
\end{aligned}
\end{equation}
where $\mathcal{L}(G, G')$ is the structural loss measured as the mean squared
error between original and reconstructed edge weights. The tolerances are
calibrated empirically from baseline measurements on the full graph, as
described in Section~5.

\subsection{Illustrative Example}

To illustrate concretely how unconstrained aggregation can harm EDI, consider
four users --- $u_1 \in \mathcal{G}_F$ and $u_2, u_3, u_4 \in \mathcal{G}_M$
--- and three items, with ratings as shown in Table~\ref{tab:toy}. This 1F:3M ratio mirrors the gender imbalance of the MovieLens 100k dataset
(29\% female, 71\% male), our primary evaluation benchmark.

Ranking items by average score yields $i_2 \succ i_3 \succ i_1$, because $i_1$
is rated 5 by $u_1$ but only 1 and 2 by two male users, pulling its mean down
to 2.67. The top-2 recommendation $R = \{i_2, i_3\}$ exposes a severe fairness
problem. Computing the EDI metrics on $R$ (with $\theta = 4$):
\begin{align*}
  \mathrm{utility}_F(R) &= \tfrac{1}{2}(w(u_1,i_2)+w(u_1,i_3))
                         = \tfrac{1}{2}(2+0) = 1.0, \\
  \mathrm{utility}_M(R) &= \tfrac{1}{6}\sum_{j=2}^{4}(w(u_j,i_2)+w(u_j,i_3))
                         = \tfrac{26}{6} = 4.33, \\
  \Delta E(R) &= |1.0 - 4.33| = 3.33.
\end{align*}
Moreover, $u_1$ finds no item in $R$ with $w(u_1,i) \geq 4$, so
$\mathrm{inclusion}_F(R) = 0 < \alpha_F = 0.25$, while
$\mathrm{inclusion}_M(R) = 0.83 > \alpha_M = 0.75$: the female user is
entirely excluded from the relevant items of the top-2 list.

Had $i_1$ been included instead --- $R' = \{i_1, i_2\}$ --- we would obtain
$\Delta E(R') = 0.67$ and $\mathrm{inclusion}_F(R') = 0.5 \geq \alpha_F$: a
substantially fairer outcome. Our EDI-constrained coarsening detects such
distortions at the structural level: any merge of user nodes that would lead
to recommending $R$ over $R'$ is rejected, because it increases $\Delta E$
beyond the tolerance $\varepsilon_E$.

\begin{table}[H]
\centering
\caption{Toy example: ratings (-- denotes an unobserved entry).}
\label{tab:toy}
\begin{tabular}{lcccc}
\hline
User & Group & $i_1$ & $i_2$ & $i_3$ \\
\hline
$u_1$ & F & 5  & 2  & --  \\
$u_2$ & M & 1  & 5  & 3   \\
$u_3$ & M & -- & 4  & 5   \\
$u_4$ & M & 2  & 5  & 4   \\
\hline
\end{tabular}
\end{table}

\section{Experimentation}\label{sec:experimentation}
\subsection{Datasets}

We evaluate our approach across five datasets spanning four domains,
covering both user-side and item-side equity settings.

\paragraph{MovieLens 100k and 1M~\cite{harper2015movielens}.}
MovieLens 100k contains 100,000 ratings from 943 users on 1,682 movies
(scale 1--5). Gender is the sensitive attribute on users: $\mathcal{G}_F$
(273 female, 29\%) and $\mathcal{G}_M$ (670 male, 71\%). MovieLens 1M
extends this to 6,040 users, 3,706 movies, and 1,000,209 ratings, used
for scalability validation. We set $k \in \{5, 10, 20\}$.

\paragraph{libimseti.cz.}
A Czech social dating network where users rate other users on attractiveness.
The sensitive attribute is the gender of the rating user (user-side equity).
We set $k \in \{5, 10, 20\}$.

\paragraph{Rate My Professors.}
An educational rating platform where students rate professors. The sensitive
attribute is the gender of the professor (item-side equity). We set
$k \in \{5, 10, 20\}$.

\paragraph{OpenAlex (AI/ML/CS, 2018--2023).}
An academic authorship graph built from the OpenAlex API,
comprising 99 venues, 904 authors (171 F / 733 M, $\alpha_F \approx 19\%$),
and 2,124 ratings (co-authorship counts). Gender is inferred from author
first names. The sensitive attribute is on items (item-side equity).
We set $k \in \{5, 10, 20\}$.

\begin{table}[H]
\centering
\caption{Overview of the five datasets.}
\label{tab:datasets-overview}
\begin{tabular}{lccccccc}
\hline
Dataset & Domain & $|U|$ & $|I|$ & $|E|$ & Scale & Equity & $\alpha_F$ \\
\hline
MovieLens 100k    & Movies       & 943   & 1,682  & 100,000    & 1--5  & user-side & 29\% \\
MovieLens 1M      & Movies       & 6,040 & 3,706  & 1,000,209  & 1--5  & user-side & 28\% \\
libimseti.cz      & Dating       & 1,000 & 32,677 & 97,707     & 1--10 & user-side & 50\% \\
Rate My Prof.     & Academic     & 1,981 & 59,066 & 150,800    & 1--5  & item-side & 51\%* \\
OpenAlex          & Publications & 99    & 904    & 2,124      & 1--5  & item-side & 19\% \\
\hline
\end{tabular}
\vspace{2pt}
\begin{footnotesize}
* $\alpha_F$ = proportion of women among the items (recommended
professors), not among users.
\end{footnotesize}
\end{table}

\begin{figure}[H]
\centering
\includegraphics[width=\textwidth]{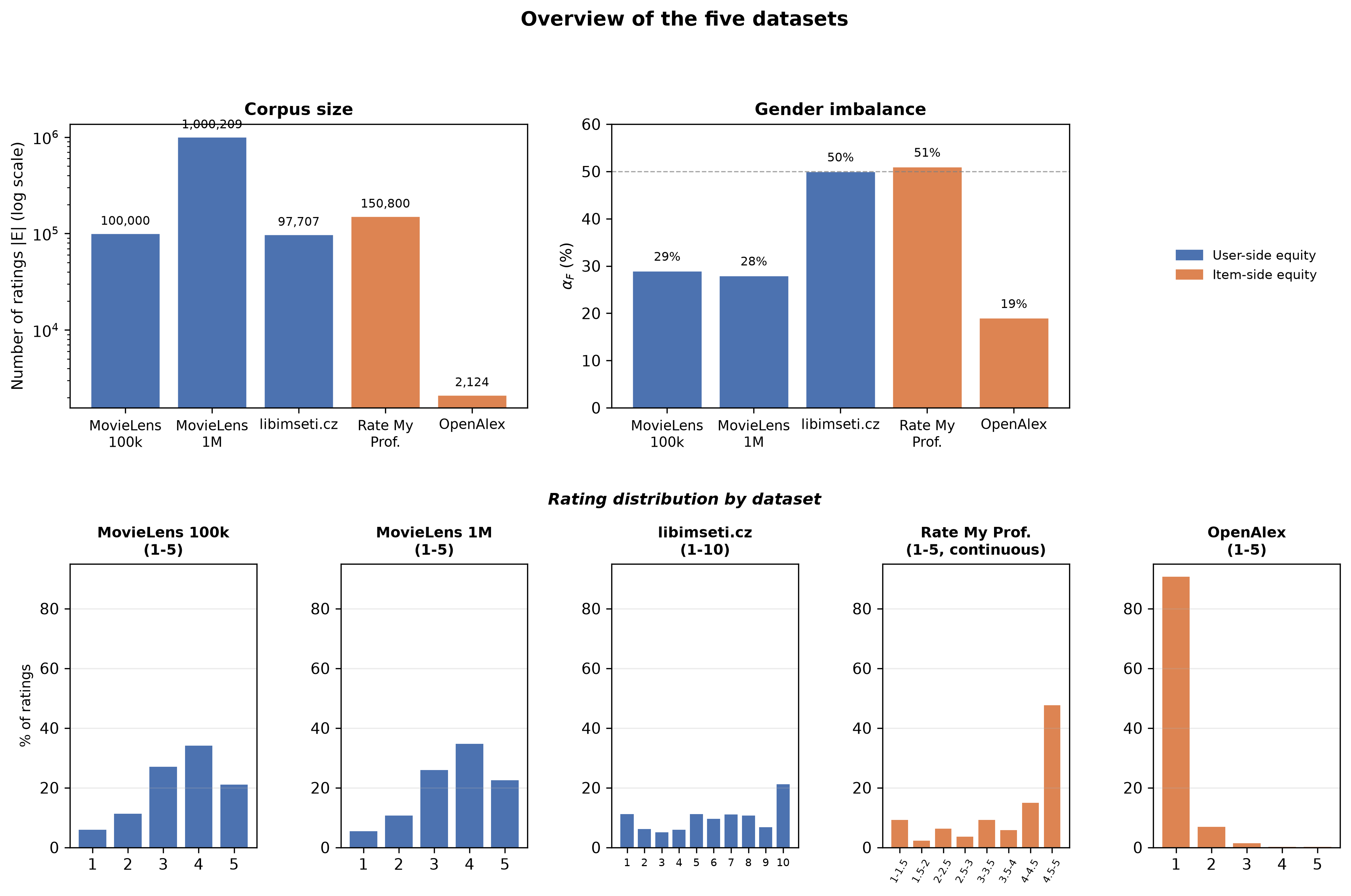}
\caption{Overview of the five datasets: corpus size (log scale) and
gender imbalance $\alpha_F$ by equity side (top), and rating distribution
for each dataset (bottom). For Rate My Professors, the rating is a
continuous (course, professor) average on 1--5, grouped into eight bins
of width 0.5.}
\label{fig:datasets-overview}
\end{figure}

\subsection{Baselines}

We compare our EDI-constrained summarization against five baselines:
(1)~\textbf{Average Score}, which ranks items by their mean rating;
(2)~\textbf{Borda Count}, which converts each user's ratings into ranks
and sums them; (3)~\textbf{Weighted Borda}, which normalizes Borda scores
by group size to compensate for demographic imbalances; (4)~\textbf{Condorcet},
which ranks items by pairwise majority preference; and (5)~\textbf{Fair Re-rank},
a post-processing baseline that greedily re-ranks the top-40 Borda candidates
to maximize the sum of group inclusion scores while minimizing the equity gap.
None of the first four rules explicitly controls EDI during aggregation;
Fair Re-rank addresses EDI as a post-hoc correction.

\subsection{Evaluation Protocol}

We evaluate all methods on the EDI metrics defined in Section~4.1.
For user-side equity datasets (MovieLens, libimseti), we report $\Delta E$,
ILD, and group inclusion scores ($\mathrm{inc}_F$, $\mathrm{inc}_M$), with
relevance threshold $\theta \in \{3.5, 4.0\}$ for MovieLens and
$\theta = 7.0$ for libimseti --- the semantic equivalent of the
MovieLens threshold on the platform's 1--10 scale, rather than a strict
linear rescaling. For item-side equity datasets (Rate My Professors,
OpenAlex), we report $\Delta E$, ILD, and $\mathrm{frac}_F$ — the fraction
of female items in the top-$k$ list, with $\theta = 4.0$ for Rate My
Professors (same 1--5 scale as MovieLens) and $\theta = 1.0$ for OpenAlex,
a minimal threshold justified by the strong sparsity of the rating matrix
(rating $= \min(\text{papers}, 5)$; most authors have only one paper per
venue, see Figure~\ref{fig:datasets-overview}).
Degradation tolerances are set to $\varepsilon_E = 0.1$, $\varepsilon_D = 0.05$,
and $\varepsilon_I = 0.05$, calibrated from baseline measurements on the full
graph. We set $k \in \{5, 10, 20\}$ for all datasets, and report the
compression ratio $|U'|/|U|$ alongside all EDI metrics.

\subsection{Results}

Tables~\ref{tab:k5}--\ref{tab:k20} report the EDI metrics and compression
ratio for all methods at $k \in \{5, 10, 20\}$ and $\theta = 4.0$.

\begin{table}[h!]
\centering
\caption{EDI metrics at $k=5$, $\theta=4.0$. \textbf{Bold}: best value
per column. $\alpha_F=0.29$, $\alpha_M=0.71$.}
\label{tab:k5}
\begin{tabular}{lccccc}
\hline
Method & $\Delta E\downarrow$ & ILD$\uparrow$ & $\mathrm{inc}_F\uparrow$
       & $\mathrm{inc}_M\uparrow$ & Ratio$\downarrow$ \\
\hline
Average Score & \textbf{0.008} & \textbf{1.000} & 0.001 & 0.002 & -- \\
Borda         & 0.491 & 0.303 & 0.337 & \textbf{0.453} & -- \\
Weighted Borda & 0.491 & 0.303 & 0.337 & \textbf{0.453} & -- \\
Condorcet     & 0.478 & 0.340 & 0.333 & 0.446 & -- \\
Fair Re-rank  & 0.151 & 0.369 & 0.297 & 0.333 & -- \\
AURORA          & 0.439 & 0.330 & \textbf{0.346} & 0.450 & \textbf{0.499} \\
\hline
$\alpha_g$ target & -- & -- & 0.29 & 0.71 & -- \\
\hline
\end{tabular}
\end{table}

\begin{table}[h!]
\centering
\caption{EDI metrics at $k=10$, $\theta=4.0$. \textbf{Bold}: best value
per column. $\alpha_F=0.29$, $\alpha_M=0.71$.}
\label{tab:k10}
\begin{tabular}{lccccc}
\hline
Method & $\Delta E\downarrow$ & ILD$\uparrow$ & $\mathrm{inc}_F\uparrow$
       & $\mathrm{inc}_M\uparrow$ & Ratio$\downarrow$ \\
\hline
Average Score & \textbf{0.007} & \textbf{1.000} & 0.001 & 0.002 & -- \\
Borda         & 0.452 & 0.349 & 0.287 & 0.393 & -- \\
Weighted Borda & 0.425 & 0.354 & 0.293 & 0.392 & -- \\
Condorcet     & 0.446 & 0.356 & 0.292 & 0.394 & -- \\
Fair Re-rank  & 0.162 & 0.429 & 0.271 & 0.313 & -- \\
AURORA          & 0.423 & 0.391 & \textbf{0.307} & \textbf{0.405} & \textbf{0.499} \\
\hline
$\alpha_g$ target & -- & -- & 0.29 & 0.71 & -- \\
\hline
\end{tabular}
\end{table}

\begin{table}[h!]
\centering
\caption{EDI metrics at $k=20$, $\theta=4.0$. \textbf{Bold}: best value
per column. $\alpha_F=0.29$, $\alpha_M=0.71$.}
\label{tab:k20}
\begin{tabular}{lccccc}
\hline
Method & $\Delta E\downarrow$ & ILD$\uparrow$ & $\mathrm{inc}_F\uparrow$
       & $\mathrm{inc}_M\uparrow$ & Ratio$\downarrow$ \\
\hline
Average Score & \textbf{0.056} & \textbf{0.956} & 0.034 & 0.045 & -- \\
Borda         & 0.424 & 0.378 & 0.246 & 0.339 & -- \\
Weighted Borda & 0.388 & 0.376 & 0.250 & 0.336 & -- \\
Condorcet     & 0.441 & 0.380 & 0.244 & 0.340 & -- \\
Fair Re-rank  & 0.240 & 0.427 & 0.258 & 0.310 & -- \\
AURORA          & 0.285 & 0.462 & \textbf{0.277} & \textbf{0.344}
              & \textbf{0.499} \\
\hline
$\alpha_g$ target & -- & -- & 0.29 & 0.71 & -- \\
\hline
\end{tabular}
\end{table}

Several observations emerge from these results. First, \textbf{Average Score}
achieves the lowest $\Delta E$ and highest ILD at all $k$ values, but this
is a degenerate outcome: it recommends identical items to all users regardless
of group, producing near-zero inclusion for both groups. Second,
\textbf{Fair Re-rank} consistently achieves the best $\Delta E$ among
competitive methods across all $k$, confirming that targeted post-processing
can reduce the equity gap effectively. Third, \textbf{our method} beats all
three classical voting rules (Borda, Weighted Borda, Condorcet) on $\Delta
E$ at every $k$, though it does not close the gap with Fair Re-rank; it
also achieves the best inclusion score for at least one group at every
$k$. At $k=10$, AURORA and Condorcet each improve all four metrics over
Borda simultaneously; Weighted Borda falls just short on
$\mathrm{inc}_M$, and Fair Re-rank, despite its large $\Delta E$
advantage, trails Borda on both inclusion criteria. At $k=20$ AURORA
improves $\Delta E$, ILD, $\mathrm{inc}_F$, and $\mathrm{inc}_M$ over
Borda, Condorcet, and Weighted Borda simultaneously, without directly
reweighting scores by group size the way Weighted Borda does. On ILD specifically, AURORA
outperforms Borda and Condorcet at every $k$, with the single exception of
$k=5$, where Condorcet is marginally higher (0.340 vs.\ 0.330); against
Fair Re-rank, AURORA's advantage only materializes at $k=20$ (0.462 vs.\
0.427), while Fair Re-rank retains a higher ILD at $k=5$ and $k=10$. Finally,
the compression ratio of our method is governed by the merge budget rather
than by $k$: with a budget expressed as a fixed fraction of $|U|$ (see
Section~4.2), the graph is reduced to approximately 50\% of its original
size (471 supernodes) at every $k \in \{5,10,20\}$, a design choice we
validate through a robustness sweep in Section~\ref{sec:robustness}.

\begin{figure}[h!]
\centering
\includegraphics[width=\textwidth]{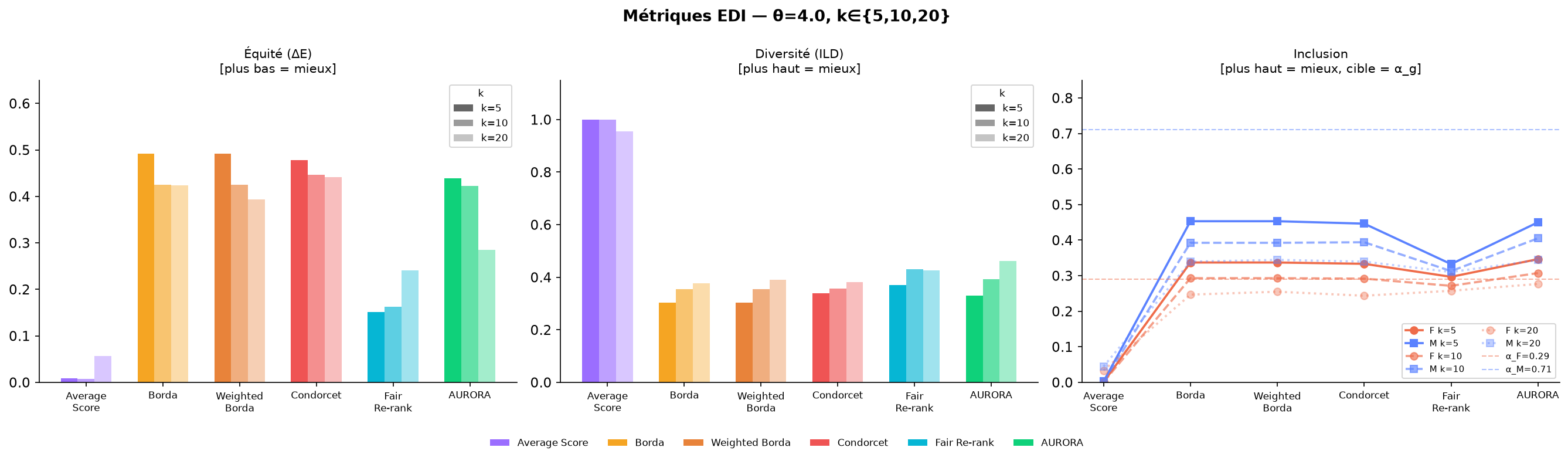}
\caption{EDI metrics for all baselines at $k \in \{5, 10, 20\}$, $\theta = 4.0$.
Lower $\Delta E$ is better (equity); higher ILD is better (diversity);
higher inclusion is better, with dashed lines marking the proportional
targets $\alpha_F=0.29$ and $\alpha_M=0.71$.}
\label{fig:baselines}
\end{figure}

\begin{table}[h!]
\caption{Average execution time (seconds) per method and list size $k$,
         measured on MovieLens 100k ($\theta = 4.0$, averaged over 3 runs).}
\label{tab:runtime}
\centering
\begin{tabular}{lrrr}
\hline
\textbf{Method} & \textbf{k = 5} & \textbf{k = 10} & \textbf{k = 20} \\
\hline
Average Score  &   0.005\,s &   0.004\,s &   0.004\,s \\
Borda          &   0.720\,s &   0.677\,s &   0.666\,s \\
Weighted Borda & ${\sim}0.72$ s & ${\sim}0.68$ s & ${\sim}0.67$ s \\
Condorcet      &   0.149\,s &   0.129\,s &   0.126\,s \\
Fair Re-rank   &   0.938\,s &   0.928\,s &   0.926\,s \\
\textbf{AURORA}  & \textbf{3.793\,s} & 3.761\,s &   4.042\,s \\
\hline
\end{tabular}
\end{table}

\paragraph{Computational efficiency.}
Table~\ref{tab:runtime} reports execution times per method. All baselines complete in
under one second. Because the merge budget is expressed as a fixed
fraction of $|U|$ rather than a fixed absolute count, the number of
candidate merges the algorithm evaluates no longer depends on $k$, and
AURORA's runtime is stable across $k \in \{5, 10, 20\}$ (3.8--4.0\,s on
MovieLens 100k). This is roughly $4$--$6\times$ slower than Fair Re-rank
and $5$--$6\times$ slower than Borda, but remains on the order of a
few seconds even as the dataset grows (Section~\ref{sec:scalability}), which we consider
acceptable given that coarsening is typically performed offline.

\subsection{Multi-Domain Evaluation}

Tables~\ref{tab:libimseti}--\ref{tab:oalex} extend the evaluation to the
three additional datasets. On \textbf{libimseti} (Table~\ref{tab:libimseti},
$\theta=7.0$ on the platform's 1--10 rating scale), Average Score achieves
both the lowest $\Delta E$ (0.016) and the highest ILD (0.867), again by
recommending the same globally popular profiles to every rater regardless of
group. Borda-based methods and AURORA all produce equity gaps above 1.2,
whereas Fair Re-rank (0.139) is a clear outlier among the non-degenerate
methods. AURORA achieves the best ILD among all methods that produce
non-trivial inclusion (0.761, ahead of Borda, Weighted Borda, and Condorcet,
which range from 0.716 to 0.746), but does not reduce $\Delta E$ relative to Borda.
We analyze this limitation in detail in
Section~\ref{sec:libimseti-limitation}: unlike the other three
datasets, libimseti is the only corpus where the sensitive attribute is
present on \emph{both} sides of the bipartite graph, and we show that the
persistently high $\Delta E$ reflects a genuine difference in rating
behavior between groups rather than an algorithmic shortcoming that
compression could fix. A robustness sweep over the merge budget (30\%--60\%
of $|U|$) confirms that this limitation is not a matter of under- or
over-compression: $\Delta E$ remains high across the whole range without
approaching the reduction seen on the other three datasets. This holds at
$k=5$ and $k=20$ too, except that at $k=5$ Fair Re-rank alone dominates
on both axes ($\Delta E=0.0004$, ILD$=0.762$), while at $k=20$ its edge
shrinks sharply ($\Delta E=0.315$) as AURORA takes the diversity lead
(ILD$=0.781$).

On \textbf{Rate My Professors} (Table~\ref{tab:rmp}), AURORA achieves the
second-highest intra-list diversity (ILD\,$=$\,0.808, behind only Average
Score's 0.982) together with the highest female professor representation
($\mathrm{frac}_F = 0.60$) among all methods. This comes at a moderate cost
on equity: AURORA's $\Delta E$ (0.112) is higher than Borda's (0.065) and
further from the near-zero gap achieved by Average Score, Weighted Borda,
and Fair Re-rank (all 0.000), so the method should be read as trading a
real amount of equity for a large gain in diversity and representation on
this dataset. Fair Re-rank raises $\mathrm{frac}_F$ to 0.50 but at the
cost of dramatically lower diversity (ILD\,$=$\,0.276), confirming that
post-hoc re-ranking cannot jointly optimize diversity and item-side
inclusion the way structural coarsening does. This trade-off shifts with
$k$: at $k=5$, AURORA has the only non-zero $\Delta E$ (0.042) but by
far the best representation ($\mathrm{frac}_F=0.80$); at $k=20$, Borda
instead has the worst $\Delta E$ (0.139), which AURORA (0.078) beats
while keeping the best ILD (0.818).

On \textbf{OpenAlex} (Table~\ref{tab:oalex}), the item-side imbalance is
stark: Borda, Weighted Borda, and Condorcet all recommend \emph{zero} female
authors ($\mathrm{frac}_F = 0.00$) despite women representing 19\% of the
author pool. Average Score and Fair Re-rank both reach $\mathrm{frac}_F = 0.20$,
while our method achieves the lowest equity gap ($\Delta E = 0.031$, a 56\%
improvement over Average Score's 0.071) while maintaining comparable female
representation ($\mathrm{frac}_F = 0.194$). Here, unlike on RMP, AURORA
improves equity \emph{and} maintains diversity simultaneously, confirming
that EDI-constrained graph coarsening generalizes meaningfully to item-side
equity settings, though its benefits are domain-dependent rather than
uniform: the bias stays severe at all three $k$ tested — at $k=5$ even
Average Score fails ($\Delta E=1.23$) for lack of candidates to dilute
it, and at $k=20$ Borda's bias only eases slightly ($\Delta E=0.907$,
5\% female authors); AURORA keeps the lowest $\Delta E$ throughout
(0.058 at $k=5$/20, 0.031 at $k=10$).

\begin{table}[h!]
\caption{EDI metrics on libimseti.cz ($k=10$, user-side equity, $\theta=7.0$
on the platform's 1--10 scale). Bold: best per column.}
\label{tab:libimseti}
\centering
\begin{tabular}{lcccc}
\hline
Method & $\Delta E \downarrow$ & ILD $\uparrow$ &
$\mathrm{inc}_F \uparrow$ & $\mathrm{inc}_M \uparrow$ \\
\hline
Average Score  & \textbf{0.016} & \textbf{0.867} & 0.002 & 0.000 \\
Borda          & 1.301 & 0.716 & 0.143 & \textbf{0.024} \\
Weighted Borda & 1.301 & 0.716 & 0.143 & \textbf{0.024} \\
Condorcet      & 1.296 & 0.746 & \textbf{0.157} & \textbf{0.024} \\
Fair Re-rank   & 0.139 & 0.716 & 0.038 & \textbf{0.024} \\
AURORA           & 1.232 & 0.761 & 0.148 & \textbf{0.024} \\
\hline
\end{tabular}
\end{table}

\subsection{Why $\Delta E$ Remains High on libimseti}
\label{sec:libimseti-limitation}

On the three other datasets, the sensitive attribute lives on only one side
of the bipartite graph: gender is a property of users on MovieLens, and a
property of items on Rate My Professors and OpenAlex. libimseti is
different: both the raters and the rated profiles carry a gender label,
because the items being recommended are themselves people. To understand
why AURORA, like the voting rules, does not reduce $\Delta E$ on this
dataset, we cross-tabulate the mean rating by the gender pair (rater,
rated profile) across the full 17.3M-edge libimseti graph (12{,}258{,}052
ratings with gender known on both sides), shown in
Table~\ref{tab:libimseti-crosstab}.

\begin{table}[h!]
\caption{Mean rating by rater--profile gender pair on libimseti.cz
(1--10 scale, full graph).}
\label{tab:libimseti-crosstab}
\centering
\begin{tabular}{lcc}
\hline
Rater $\to$ Rated & Mean rating & \# ratings \\
\hline
F $\to$ M & 6.92 & 7{,}099{,}688 \\
M $\to$ F & 5.48 & 3{,}232{,}064 \\
F $\to$ F & 5.14 & 1{,}243{,}590 \\
M $\to$ M & 4.46 &   682{,}710 \\
\hline
\end{tabular}
\end{table}

The gap between the highest mean rating (F$\to$M, 6.92) and the lowest
(M$\to$M, 4.46) is large and consistent across more than 12 million
ratings: it is a structural property of the data, not statistical noise.
A single collective ranking, shared by every rater regardless of group,
cannot simultaneously satisfy raters whose underlying rating distributions
differ this much. This explains two observations from
Table~\ref{tab:libimseti}: first, $\Delta E$ is elevated for the voting
methods (Borda, Weighted Borda, Condorcet) and for AURORA alike --- Average
Score and Fair Re-rank are the exceptions, but only because neither
produces a ranking that differentiates between groups in the first place;
second, the EDI-preservation constraint on the coarsening process has
essentially no effect here, since the top-$k$ list is already dominated by
profiles that are highly rated across the board, and merging users of the
same gender cannot transfer information about a group's preferences for
items it rates comparatively lower.

This result is an empirical illustration of an argument made by Yao and
Huang~\cite{yao_new_2017}: demographic parity ($\Delta E \approx 0$) is not
always an appropriate fairness target when the preferences of different
groups genuinely diverge. On MovieLens or Rate My Professors, nothing
about the domain suggests that a user's gender should influence how much
they enjoy a film or how highly they rate a professor's teaching, so a
persistent $\Delta E$ gap there is plausibly attributable to the
aggregation rule. On libimseti, by contrast, the sensitive attribute
enters the preference signal itself through the structure of romantic and
social attraction, so a high $\Delta E$ may reflect a genuine difference
in preference rather than an unfair outcome to correct. We verified that
this is not a matter of under-calibration: across the same merge-budget
sweep described in Section~\ref{sec:robustness} (30\%--60\% of $|U|$),
$\Delta E$ remains elevated throughout, while AURORA's diversity advantage
over the voting methods persists regardless of budget. We view the
elevated $\Delta E$ as a genuine limitation of a single-ranking
aggregation framework on domains with reciprocal, attribute-dependent
preferences, rather than a defect specific to our coarsening method.

\begin{table}[h!]
\caption{EDI metrics on Rate My Professors ($k=10$, item-side equity).
$\mathrm{frac}_F$ = fraction of female professors in top-$k$.
Bold: best per column.}
\label{tab:rmp}
\centering
\begin{tabular}{lccc}
\hline
Method & $\Delta E \downarrow$ & ILD $\uparrow$ &
$\mathrm{frac}_F \uparrow$ \\
\hline
Average Score  & \textbf{0.000} & \textbf{0.982} & 0.40 \\
Borda          & 0.065 & 0.267 & 0.50 \\
Weighted Borda & \textbf{0.000} & 0.245 & 0.20 \\
Condorcet      & 0.076 & 0.253 & 0.30 \\
Fair Re-rank   & \textbf{0.000} & 0.276 & 0.50 \\
AURORA           & 0.112 & 0.808 & \textbf{0.60} \\
\hline
\end{tabular}
\end{table}
\FloatBarrier
\begin{table}[h!]
\caption{EDI metrics on OpenAlex AI/ML/CS 2018--2023 ($k=10$, item-side equity;
$\alpha_F \approx 19\%$). Bold: best per column.}
\label{tab:oalex}
\centering
\begin{tabular}{lccc}
\hline
Method & $\Delta E \downarrow$ & ILD $\uparrow$ &
$\mathrm{frac}_F \uparrow$ \\
\hline
Average Score  & 0.071 & \textbf{0.356} & \textbf{0.200} \\
Borda          & 1.233 & 0.000 & 0.000 \\
Weighted Borda & 1.233 & 0.000 & 0.000 \\
Condorcet      & 1.233 & 0.000 & 0.000 \\
Fair Re-rank   & 0.071 & \textbf{0.356} & \textbf{0.200} \\
AURORA           & \textbf{0.031} & 0.345 & 0.194 \\
\hline
\end{tabular}
\end{table}
\FloatBarrier

\subsection{Robustness to the Merge Budget}
\label{sec:robustness}

The merge budget (Section~\ref{sec:methodology}) is set to 50\% of $|U|$
throughout the main experiments. To check that this choice is not a
narrow, cherry-picked setting, we re-run AURORA at four budgets --- 30\%,
42.4\%, 50\%, and 60\% of $|U|$ --- on MovieLens 1M and Rate My
Professors, the two datasets where the effect of compression is most
visible. Table~\ref{tab:robustness} reports $\Delta E$ and ILD at each
budget.

\begin{table}[h!]
\caption{AURORA's $\Delta E$ and ILD across four merge budgets.
MovieLens 1M at $k=20$; Rate My Professors at $k=10$.}
\label{tab:robustness}
\centering
\begin{tabular}{lcccc}
\hline
Budget ($\%$ of $|U|$) & \multicolumn{2}{c}{MovieLens 1M ($k=20$)} &
\multicolumn{2}{c}{Rate My Professors} \\
 & $\Delta E \downarrow$ & ILD $\uparrow$ & $\Delta E \downarrow$ &
ILD $\uparrow$ \\
\hline
30\%   & 0.332 & 0.456 & 0.112 & 0.745 \\
42.4\% & 0.277 & 0.456 & 0.071 & 0.788 \\
50\%   & 0.277 & 0.456 & 0.112 & 0.808 \\
60\%   & 0.185 & 0.470 & 0.112 & 0.767 \\
\hline
\end{tabular}
\end{table}

On MovieLens~1M, AURORA beats Borda, Weighted Borda, and Condorcet on both
$\Delta E$ and ILD at every budget in this range, and both metrics improve
slightly as the budget grows --- the method is robust to the exact choice
of budget, and 50\% is a reasonable, round value within a range that
performs consistently well. On Rate My Professors, ILD is stable and
consistently high (0.75--0.81) across the range, confirming that the
diversity gain is not an artefact of the specific 50\% budget; $\Delta E$,
by contrast, does not vary smoothly with the budget and is not
monotonically improved by more or less compression, which we attribute to
the coarse granularity of a $k=10$ ranking (only a limited number of
distinct top-$k$ configurations are reachable as merges accumulate). Borda
remains close to $\Delta E \approx 0.065$ throughout this range on Rate My
Professors, so the equity trade-off described in Section~5.4 holds
regardless of the exact budget chosen. We did not observe a budget within
30--60\% that closes this gap while preserving AURORA's diversity
advantage, which we leave as an open question for future calibration
work (Section~6).

Are the MovieLens 100k results at $k=10$ (Table~2) stable under
resampling of the user set? Subsampling without replacement (90\% of
$|U|$, 30 repetitions) confirms they are: AURORA beats Borda and
Condorcet on $\Delta E$ in 28/30 draws and on ILD in 30/30 draws
(Weighted Borda: 24/30 and 30/30), with low variance on both metrics
($\Delta E = 0.417 \pm 0.035$, ILD $= 0.391 \pm 0.004$) --- AURORA's
advantage over the classical voting rules is not an artefact of the
single draw reported in Table~2. Fair Re-rank, by contrast, stays ahead
of AURORA on both metrics in all 30 draws without exception, confirming
that its advantage on this dataset is likewise robust.

\subsection*{Cross-Domain Summary}

Table~\ref{tab:synthesis} summarizes the best-performing method per
metric and dataset at $k=10$ (or the reported $k$ for single-$k$
datasets), excluding Average Score where it is degenerate. AURORA leads
on diversity on MovieLens 100k and on both diversity and representation
on the item-side datasets (RMP, OpenAlex), and it beats every classical
voting rule on $\Delta E$ across all five datasets. It is not, however,
the strongest method everywhere: Fair Re-rank achieves the best $\Delta
E$ among non-degenerate methods on MovieLens 100k and, notably, both the
best $\Delta E$ \emph{and} the best ILD on MovieLens 1M, making it the
overall strongest method on that dataset rather than AURORA. No single
method wins across all metrics and domains, which confirms that EDI
objectives are in tension and that no single mechanism --- structural or
post-hoc --- dominates uniformly across settings.

\begin{table}[h!]
\caption{Cross-domain summary: best method per metric at $k=10$ (or the
single reported $k$ for libimseti, RMP, OpenAlex).}
\label{tab:synthesis}
\centering
\footnotesize
\resizebox{\textwidth}{!}{%
\begin{tabular}{@{}lllll@{}}
\hline
Dataset & Best $\Delta E$ & Best ILD & Best inc/frac$_F$ &
Overall \\
\hline
ML 100k  & Fair Re-rank & Fair Re-rank & \textbf{AURORA} & \textbf{AURORA} ($k{=}20$) \\
ML 1M    & Fair Re-rank & Fair Re-rank & Fair Re-rank & Fair Re-rank \\
libimseti & Fair Re-rank & \textbf{AURORA} & Condorcet & None (persistent gap) \\
RMP      & Avg Score / W.Borda / F.R. & Avg Score & \textbf{AURORA} & Trade-off: AURORA (div.) / F.R. (eq.) \\
OpenAlex & \textbf{AURORA} & Avg Score / F.R. & Avg Score / F.R. & \textbf{AURORA} \\
\hline
\end{tabular}%
}
\par\smallskip
\footnotesize $^*$Average Score is excluded above wherever it is
degenerate on inclusion (near-zero $\mathrm{inc}_F$/$\mathrm{inc}_M$): on
the three user-side datasets --- ML 100k, ML 1M, and libimseti alike ---
recommending the same list to every user regardless of group leaves it
with almost no group-specific relevant items. On the two item-side
datasets (RMP, OpenAlex), $\mathrm{frac}_F$ is not similarly degenerate for
Average Score, so it is not excluded there. F.R.\ = Fair Re-rank; AURORA
is a close third on OpenAlex ILD and $\mathrm{frac}_F$.
\end{table}

\subsection{Scalability}
\label{sec:scalability}
To assess whether EDI-constrained coarsening scales beyond MovieLens
100k, we evaluate all methods on MovieLens 1M (6,040 users, 3,706
items, 1,000,209 ratings) at $k \in \{5, 10, 20\}$ and $\theta = 4.0$.
Table~\ref{tab:1m} reports results at $k = 10$.
\begin{table}[h!]
\caption{EDI metrics on MovieLens 1M ($k=10$, $\theta=4.0$).
Bold: best per column.}
\label{tab:1m}
\centering
\begin{tabular}{lcccc}
\hline
Method & $\Delta E \downarrow$ & ILD $\uparrow$ &
$\mathrm{inc}_F \uparrow$ & $\mathrm{inc}_M \uparrow$ \\
\hline
Average Score  & \textbf{0.000} & \textbf{1.000} & 0.000 & 0.000 \\
Borda          & 0.482 & 0.418 & 0.305 & 0.409 \\
Weighted Borda & 0.414 & 0.426 & 0.311 & 0.400 \\
Condorcet      & 0.487 & 0.408 & 0.305 & \textbf{0.412} \\
Fair Re-rank   & 0.019 & 0.452 & \textbf{0.336} & 0.343 \\
AURORA           & 0.214 & 0.440 & 0.321 & 0.370 \\
\hline
\end{tabular}
\end{table}

\begin{figure}[H]
\centering
 \includegraphics[width=\textwidth]{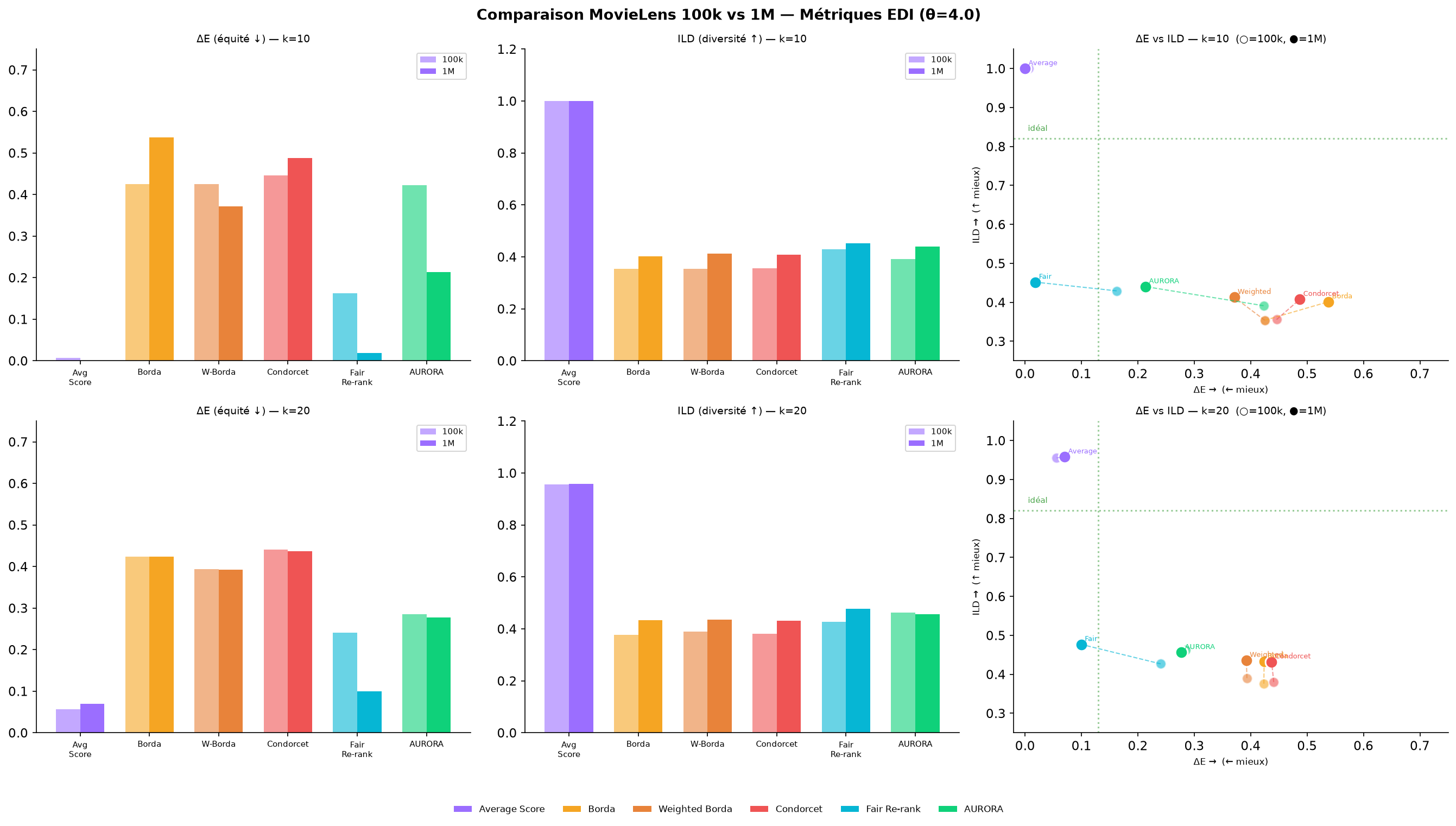}
\caption{Side-by-side comparison of $\Delta E$ and ILD between MovieLens
100k and 1M, for all six methods at $k \in \{10, 20\}$ (top row: $k=10$;
bottom row: $k=20$). Lighter bars denote MovieLens 100k, darker bars
denote MovieLens 1M.}
\label{fig:comparison1m}
\end{figure}

The relative ordering among the classical voting rules is broadly consistent
with MovieLens 100k: Weighted Borda improves over Borda on $\Delta E$ (0.482
$\to$ 0.414 at $k=10$), and AURORA beats all three voting rules (Borda, Weighted
Borda, Condorcet) on both $\Delta E$ and ILD at every $k \in \{5,10,20\}$.
However, unlike on MovieLens 100k, Fair Re-rank dominates AURORA on
\emph{both} $\Delta E$ and ILD at every $k$ on this larger dataset (e.g.\ at
$k=10$: $\Delta E=0.019$ and ILD$=0.452$ for Fair Re-rank, versus 0.214 and
0.440 for AURORA). We attribute this to the larger user base making the
post-hoc re-ranking baseline more effective: with 6,040 users to draw
candidates from, Fair Re-rank's greedy re-ranking of the top-40 Borda
candidates has considerably more room to satisfy both groups than on the
943-user MovieLens 100k graph. AURORA's most distinctive result on 1M is at
$k=20$, where it remains competitive on $\mathrm{inc}_F$ (0.294, within
0.001 of the best, Weighted Borda at 0.295, and ahead of Fair Re-rank at
0.291), suggesting that its advantage is concentrated in balancing inclusion across
groups rather than in minimizing $\Delta E$ or maximizing ILD outright at
this scale. Figure~\ref{fig:comparison1m} shows the side-by-side comparison of $\Delta E$ and
ILD across all methods at $k \in \{10, 20\}$.

\section{Conclusion and Future Work}
\label{sec:conclusion}

This paper addressed the problem of aggregating user preferences in a way that
explicitly preserves equity, diversity, and inclusion. We represented preferences
as a weighted attributed bipartite graph and proposed AURORA, an EDI-constrained
coarsening algorithm that embeds fairness directly into the graph structure: each
candidate merge of user nodes is accepted only if the resulting summarized graph
maintains the equity gap, intra-list diversity, and group inclusion of the original
within prescribed tolerances. This structural approach contrasts with existing
post-hoc techniques such as re-ranking, which enforce fairness after the
aggregation has already been computed.

Evaluated across five datasets spanning four domains, no single baseline
satisfies all three EDI criteria simultaneously. On MovieLens 100k, AURORA
consistently beats the classical voting rules (Borda, Weighted Borda, Condorcet)
on both equity and diversity across $k \in \{5,10,20\}$, with a merge budget
robust across a 30--60\% range; Fair Re-rank remains the most accurate on equity
alone, at some cost to diversity. This picture is dataset-dependent rather than
universal: on the larger MovieLens~1M, Fair Re-rank dominates AURORA on both
axes, which we attribute to its larger candidate pool for re-ranking; on
libimseti, where the sensitive attribute sits on both sides of the bipartite
graph, AURORA --- like the classical voting rules --- does not reduce the equity gap,
a limitation traced to a genuine difference in rating behavior between groups
rather than an algorithmic shortcoming. On the item-side datasets, AURORA shows
two distinct patterns: on Rate My Professors it trades equity for a large gain
in diversity (ILD\,$=$\,0.808) and female representation ($\mathrm{frac}_F=60\%$),
while on OpenAlex it improves equity \emph{and} maintains diversity
simultaneously, reaching the lowest equity gap observed across all five datasets
($\Delta E = 0.031$) and correcting the most extreme bias in our evaluation.

Taken together, these results support a more precise claim than "AURORA is
uniformly better": embedding EDI constraints into the coarsening structure
consistently outperforms unconstrained voting rules, and matches or improves
on post-hoc re-ranking on smaller or item-side datasets, but a sufficiently
large user base or a domain with reciprocal, attribute-dependent preferences
can favor a post-hoc approach or expose a genuine limit of any single-ranking
aggregation framework. This nuance is itself a contribution: it clarifies the
conditions under which structural, constraint-based summarization is
preferable to re-ranking, rather than asserting unconditional superiority.

Several directions remain open for future work: extending the framework to
multiple simultaneous sensitive attributes (e.g., gender combined with age or
occupation); adapting the static coarsening process to dynamic or streaming
preference graphs; deriving the tolerances $\varepsilon_E$, $\varepsilon_D$,
$\varepsilon_I$ and the merge budget from a principled criterion --- such as a
Pareto-frontier analysis or a held-out validation split --- rather than an
empirical robustness sweep; understanding why Fair Re-rank overtakes AURORA on
MovieLens~1M but not on smaller datasets, to clarify the boundary conditions
favoring each mechanism; on Rate My Professors specifically, designing a merge
criterion that accounts for the downstream item distribution rather than
rating-profile similarity alone, to narrow the equity--diversity trade-off
without sacrificing representation gains; and replacing the greedy merge
strategy with community-detection techniques guided directly by the EDI impact
of each candidate merge.

\bibliographystyle{splncs04}
\bibliography{references}
\end{document}